\documentclass[epj]{svjour}
\usepackage{epsf}
\usepackage{latexsym}
\usepackage{amssymb}
\begin{document}
\title{Vibrational States of Glassy and Crystalline Orthoterphenyl}
\titlerunning{Vibrational States of Orthoterphenyl}
\author{A.~T\"olle\inst{1} 
    \and H.~Zimmermann\inst{2}
    \and F.~Fujara\inst{3}
    \and W.~Petry\inst{4}
    \and W.~Schmidt\inst{5}
    \and H.~Schober\inst{5}
    \and J.~Wuttke\inst{4}
}                     
\authorrunning{A.~T\"olle {\it et al.}}
\institute{Fachbereich Physik, Universit\"at Dortmund, 44221 Dortmund, Germany
  \and Max-Planck-Institut f\"ur medizinische Forschung, 69120 Heidelberg, 
       Germany 
  \and Institut f\"ur Festk\"orperphysik, Technische Universit\"at Darmstadt,
       64289 Darmstadt, Germany
  \and Physik-Department E13, Technische Universit\"at M\"unchen, 85747
       Garching, Germany
  \and Institut Laue-Langevin, 38042 Grenoble, France
} 
\date{\today}
%
\abstract{
Low-frequency vibrations of glassy and crystalline
orthoterphenyl are studied by means of neutron scattering.
Phonon dispersions are measured along the main axes of a single crystal,
and the corresponding longitudinal and transversal sound velocities 
are obtained.
For glassy and polycrystalline samples, a density of vibrational states
is determined and cross-checked against other dynamic observables.
In the crystal, low-lying zone-boundary modes lead to an excess over
the Debye density of states. 
In the glass, the boson peak is located at even lower frequencies.
With increasing temperature, both glass and crystal
show anharmonicity.
} 
\maketitle
\section{Introduction} \label{Intro}

Orthoterphenyl (OTP)
has been studied for more than fourty years as
prototype of a non-associative, non-polar molecular glass former \cite{AnUb55}.
Early studies concentrated on viscosity 
\cite{AnUb55,LaUb58,GrTu67,LaUh72,CuLU73,ScKB98} 
and $\alpha$-relaxation 
\cite{WiHa71,WiHa72,Arr75,VaFl80,FyWL81,DiNa88,MeFi88,FiBH89}
as direct manifestations of the frequency-dependent glass transition,
or on slow $\beta$ relaxation \cite{JoGo70a,JoGo70b,WuNa92}.

In the past decade, a new frame has been set by mode-coupling theory
\cite{Got91,GoSj95,Cum99}
which describes the onset of structural relaxation on microscopic time scales.
Consequently, 
the fast dynamics of OTP has been studied 
by incoherent \cite{BaFK89b,PeBF91,KiBD92,WuKB93,ToSW98a}
and coherent \cite{BaFL95,ToSW97b,ToWS98b} 
inelastic neutron scattering as well as
by depolarised light scattering \cite{StPG94,SoSR95,CuLD97b}
and is found to be in good accord with asymptotic results of theory.

Closed formulations of mode-coupling theory exist only 
for very simple systems consisting of 
spherical particles \cite{BeGS84} or
mixtures thereof \cite{BoTh87},
and the generalisation to dipolar molecules requires already
intimidating formal efforts \cite{FrFG97c,ScSc97}.
Intramolecular vibrations have not yet been explicitely considered,
and therefore the fit of mode-coupling asymptotes to complex systems
like OTP remains heuristic.
Worse: theoretical studies of a hard-spheres system \cite{FrFG97b,FuGM98}
have shown that the full asymptotic behaviour is only reached 
when the relaxational time scale is separated by several orders of magnitude
from the microscopic dynamics,
whereas experimental studies are restricted to at best three decades
in time or frequency.
On this background the present communication
shall supplement our neutron scattering studies 
of relaxational dynamics in OTP 
\cite{BaFK89b,PeBF91,KiBD92,WuKB93,ToSW98a,BaFL95,ToSW97b,ToWS98b}
by an explicit investigation of vibrational states.

As other glasses, OTP possesses more low-frequency modes than expected from
the Debye model.
This excess, for obscure reasons named ``boson-peak'',
is usually taken as a characteristic feature of disordered systems.
To substantiate this interpretation,
glasses and crystals must be compared systematically, 
which so far has been done for very few systems
\cite{Lea69,GoLa81,Gom81,CrBA94,BeCM94,BeCA96,MeWP96a,DoHH97,CuBF98,TaRV98,BeCC98}.

OTP crystallises easily into a polycrystalline powder.
Some effort is needed to prevent it from crystallisation when
cooling below the melting point T$_m$=329\,K.
It forms a stable glassy structure when it is supercooled below 
the caloric glass transition temperature T$_g$=243\,K.
One particular advantage is that large single crystal specimens
of several cm$^3$ can be grown.
So we are able to compare scattering from a single crystal,
from powder-like polycrystal, and from the glass.

The paper is organised as follows:
The analysis of single crystal dispersion relations in Sec.~\ref{sect2}
enables us to fix the frequency range of acoustic modes in the ordered
state.
The vibrational density of states and the derived thermodynamic quantities 
of the glass and the polycrystal are compared with each other in 
Sec.~\ref{sect3}. 
The temperature effect on the frequency distributions in the crystal and the
glass is presented in Sec.~\ref{sect4}.

\section{Single Crystal: Phonon Dispersions} \label{sect2}

\subsection{Structural Information} \label{subsect21}

The OTP molecule (1,2-diphenyl-benzene: C$_{18}$H$_{14}$) consists of a 
central benzene ring and two lateral phenyl rings in ortho position.

The crystal structure belongs to the orthorhombic space group $P2_12_12_1$
with four molecules per unit cell and lattice parameters
$a=18.583$\,\AA, $b=6.024$\,\AA, $c=11.728$\,\AA\
at room temperature \cite{BrLe79,AiMO78}.
A sketch of the structure is given in Fig.~1 of Ref.~\cite{BrLe79}.

For steric reasons, the lateral phenyl rings are necessarily rotated
out of plane.
In addition, the overcrowding in the molecule leads to
significant bond-angle and out-of-plane distortions of the phenyl-phenyl bonds.
Such structural irregularities may explain why OTP can be undercooled
far easier than $m$- or $p$-terphenyl \cite{AnUb55}.

In the crystal, 
the angles for the out-of-plane rotation of the lateral phenyl rings 
are $\phi_1\simeq43^\circ$ and $\phi_2\simeq62^\circ$ \cite{BrLe79,AiMO78}.
For isolated molecules,
an old electron diffraction study had suggested
$\phi_1=\phi_2=90^\circ$,
but newer experiments and calculations agree that in the gas or liquid phase
$40^\circ \lesssim \phi_1, \phi_2 \lesssim 65^\circ$ \cite{BaPo85}.

\subsection{Triple-Axis Experiment} \label{subsect22}

For coherent neutron scattering,
perdeuterated OTP (C$_{18}$D$_{14}$, $>$ 99\,\% deuteration) was used.
Single crystals of high quality and considerable size (several cm$^3$) 
were grown out of hot methanol solution 
either by very slow cooling or by evaporation over several months.
They grew preferentially along the shortest axis $b$.

The phonon dispersion measurements were carried out on a specimen 
of about $1\times1\times2$\,cm$^3$ for temperatures between 100 and 310\,K. 
The experiments were performed on the cold triple-axis spectrometer IN12 
at the Institut Laue Langevin (ILL),
mostly with constant final wave vector $k_f=1.55$\,\AA$^{-1}$.

The lattice parameters as measured on IN12 at room temperature 
$a=18.53$\,\AA, $b=6.02$\,\AA, $c=11.73$\,\AA\ 
are in good agreement with the literature results from 
neutron \cite{BrLe79} and X-ray diffraction \cite{AiMO78}.
On cooling to 100\,K, $a$ and $c$ change only by a few ppm,
whereas $b$ contracts by about 3\,\%.

\subsection{Acoustic Phonons} \label{subsect23}

In Fig.~1 we present the measured phonon dispersion of 
crystalline OTP at 200\,K along the three 
main symmetry directions [100], [010], and [001].
Within the available beam time not all acoustic modes 
could be investigated.
Some optic phonons were detected as well, 
but not studied in detail;
as an example, 
the lowest optic branch in [010] direction is included in Fig.~1.

Due to the crystal symmetry modes of longitudinal and transverse character
are pairwise degenerated at the Brillouin zone boundary.
A first interesting point is that
purely acoustic modes are confined to a rather small frequency range:
crossing with optic-like branches occurs already at about 0.6\,THz. 

Lattice dynamics calculations \cite{CrBA94} agree qualitatively
with the measured acoustic dispersions.
For instance, the crossing of the longitudinal acoustic mode in [010] direction
with optic branches is predicted correctly.
Quantitatively, 
the measured acoustic dispersions are considerably steeper than calculated.
The sound velocities, determined from the initial slopes, 
are summarised in Table~\ref{tab:1};
they lie 20--60\% above the calculated values.
In all lattice directions,
longitudinal sound modes are almost twice as fast as transverse modes.

The single crystal data may also be compared with
sound velocities in the glass.
Results from Brillouin scattering 
are summarised in Table~\ref{TsoundG}.
Taking the simple arithmetic average over the three crystal axes,
the mean longitudinal and transverse sound velocities 
$\langle v_{\rm L,T} \rangle$
exceed those of the glass by about 20\,\% and 27\,\%, respectively.

\section{Glass and Polycrystal: Density of Vibrational States} \label{sect3}

\subsection{Density of States and Neutron Scattering} \label{dosdos}

In the absence of crystalline order,
it is no longer possible
to measure selected phonon modes with well-defined polarisation and
propagation vector.
For polycrystalline or amorphous samples,
the distribution of vibrational modes can be conveyed only
in form of a spectral density of states (DOS).

In principle, the DOS can be measured 
in absolute units by incoherent neutron scattering.
However, as soon as one goes beyond the simplest textbook example
of a harmonic, monoatomic, polycrystalline solid
with a simple (ideally cubic) lattice,
several difficulties arise and the very concept of a DOS becomes problematic.

In a molecular solid, 
different atoms~$j$ participate in given vibrational modes~$r$ 
with different amplitudes ${\bf e}^r_j$.
Therefore, 
atoms in non-equivalent positions have 
different vibrational densities of states $g_j(\nu)$.
In neutron scattering,
these $g_j(\nu)$ are weighted with the scattering cross sections $\sigma_j$.
In the case of protonated OTP, 
we see almost only incoherent scattering from hydrogen.
Worse, in the scattering law $S(q,\nu)$ the $g_j(\nu)$ 
of non-equivalent hydrogen atoms
are weighted with a prefactor $|{\bf e}^r_j|^2$ and 
a Debye-Waller factor that depends also on~$j$.

Only for low-frequency, long-wavelength vibrations
the mo\-le\-cules (or some structural subunits) move as rigid bodies,
the $g_j(\nu)=g(\nu)$ become the same for all~$j$,
and the ${\bf e}^r_j$ become independent of~$r$ \cite{WuKB93,CaPe75a}.
In this limit, the determination of $g(\nu)$ from a neutron scattering law
remains meaningful and feasible.

\subsection{Experiments and Data Reduction} \label{dosexp}

Vibrational spectra from glassy OTP between 160 and 245\,K 
have been analysed previously \cite{WuKB93}.
For the present comparison,
we measured incoherent scattering from the glass at 100\,K, 
and from the polycrystalline powder at 100, 200, and 300\,K
on the time-of-flight spectrometer IN6 
at the ILL with an incident wavelength of 5.1\,\AA.
If not stated otherwise,
we refer in the following to the 100\,K data.
In these experiments, protonated OTP was used.
OTP of $>$ 99\,\% purity was bought from Aldrich 
and further purified by repeated vacuum distillation.

The raw data were converted into $S(2\theta,\nu)$ and the container 
scattering subtracted.
Without interpolating to constant wavenumbers~$Q$,
we calculated the DOS directly from $S(2\theta,\nu)$,
preferentially using data from large scattering angles $2\theta$.
Multi-phonon contributions were calculated 
by repeatedly convoluting $g(\nu)$ with itself
and subtracting it from $S(2\theta,\nu)$ in an iterative procedure,
as described in detail in \cite{WuKB93}.

The so-obtained DOS shows a pronounced gap above $\nu_{\rm g}\simeq5$\,THz.
Comparing our results with model calculations \cite{Bus82},
we assign all vibrations below $\nu_{\rm g}$ to
the 16 degrees of freedom needed to describe the crystal structure.
For 16 low-lying modes in a molecule with 32 atoms,
we expect an integrated DOS
\begin{equation}
    \int_0^{\nu_{\rm g}}\!{\rm d}\nu\, g(\nu)={16\over32}=0.5\,.
\end{equation}
This condition is used to readjust the absolute scale of $g(\nu)$. 
In the Appendix, we argue that the difference between measured and 
re-normalised $g(\nu)$ is mainly an effect of multiple scattering.

\subsection{Density of States in Orhoterphenyl} \label{DOSOTP}

Fig.~2 shows the DOS of glassy and crystalline OTP at 100\,K.
Rather broad distributions are found for both phases.
In the glass a first shoulder around 1.5\,THz is followed by a second at 
3.5\,THz in accordance with results from Raman studies \cite{CrBA94,KiVP99}.
As expected, the crystal DOS is more structured, in particular in the 
low energy region. 
Distinct peaks at 0.6, 0.8, 1.1 and 1.5\,THz become apparent.
They are due to strong contributions from zone-boundary modes,
as can be seen from Fig.~1.
Compared to the glass, significant density is missing in the low 
energy region and in the range from 1.5 to 3\,THz, 
and reappears at higher frequencies around 3.5\,THz.

In order to show the low energy modes on enhanced scale,
we plot in Fig.~3 $g(\nu)/\nu^2$ 
which in the one-phonon approximation is proportional to $S(Q,\nu)$ itself.
In this representation,
the excess of the glass over the crystal becomes evident.
A well defined frequency peak appears around 0.35\,THz which is downshifted 
with respect to the first peak of the crystal at 0.6\,THz
and superposed to a long tail which is similar for glass and crystal. 
Note that the maximum of the boson peak at 0.35\,THz is located 
below the lowest acoustic zone-boundary phonons in the crystal.

The DOS of the polycrystalline sample
is in accord with the measured dispersion of the single crystal:
A small shoulder at 0.4\,THz can be attributed 
to the transverse acoustic zone-boundary phonons in $[100]$ direction,
and the main peak at 0.6\,THz corresponds 
to the transverse acoustic zone-boundary phonons in the other two lattice 
directions.
The peak at 0.8\,THz reflects the longitudinal acoustic zone-boundary 
phonon in $[001]$ direction and the transverse optic in $[010]$ direction.

We performed additional coherent scattering experiments on a deuterated 
sample, which show that the frequency of the boson peak maximum 
has no dispersion for wave numbers in the range 0.8 to 2\,\AA$^{-1}$.
Its intensity is modulated in phase with the static structure factor,
in accordance with observations in other glasses \cite{MeWP96a,BuWR96b}. 

\subsection{Mean Square Displacement} \label{mean-square}

Another dynamic observable which can be obtained from neutron scattering
is the atomic mean square displacement $\langle r^2(T) \rangle$.
Roughly speaking, $\langle r^2(T) \rangle^{3/2}$ measures the volume
to which an atom remains confined in the limit $t\rightarrow\infty$.

For a large class of model situations 
(harmonic solid, Markovian diffusion, \ldots) 
it can be obtained directly
from the Gaussian $Q$ dependence of the elastic scattering intensity
\begin{equation} \label{Sq0}
         S(Q,\nu\!=\!0)=\exp(-Q^2 \langle r(T)^2 \rangle)\,.
\end{equation}
For a harmonic solid, 
\begin{equation} \label{msd}
      \langle r(T)^2\rangle = 
         {\hbar^2 \over 6 M k_{\rm B} T } \int_0^\infty\! {\rm d}\nu\,
         \frac{g(\nu)}{\beta} {\rm coth}(\frac{\beta}{2}) 
\end{equation}
with $\beta=h\nu/k_{\rm B}T$
crosses over from zero-point oscillations 
$\langle r^2(0) \rangle$
to a linear regime
$\langle r^2(T) \rangle \propto T$ \cite{x43}.
In any real experiment,
which integrates over the elastic line with a resolution $\Delta\nu$,
one measures actually atomic displacements within finite times 
$t_\Delta\simeq 2\pi/\Delta\nu$.

Fig.~4 shows mean square displacements of OTP,
determined according to (\ref{Sq0}) 
from elastic back-scattering 
(fixed window scans on IN13, with $t_\Delta\simeq100-200\,$psec)
and from Fourier-deconvoluted time-of-flight spectra
(taking the plateau $S(Q,t_\Delta)$ with $t_\Delta\simeq5-10\,$psec
from IN6 data that were Fourier transformed and divided through the
Fourier transform of the measured resolution function).
For the glassy sample, a direct comparison can be made and shows
good agreement between IN6 and IN13.
In the polycrystalline sample,
the displacement is for all temperatures smaller than in the glass.

The lines in Fig.~4
are calculated through (\ref{msd}) from the DOS at 100\,K.
For low temperatures, the $\langle r^2(T) \rangle$ 
are in full accord with the values determined through (\ref{Sq0}).
This comparison can be seen as a cross-check between
the analysis of elastic and inelastic neutron scattering data.

Equation (\ref{msd}) gives not only the absolute value 
of~$\langle r^2(T) \rangle$,
but enables us also to read off which modes contribute most
to the atomic displacement.
To this end, we restrict the integration (\ref{msd}) to 
modes with $0<\nu<\nu'$.
The inset in Fig.~4 
shows the relative value 
$\langle r^2(\nu';T) \rangle^{1/2} / \langle r^2(\nu;T) \rangle^{1/2}$ 
for $T=100\,$K as function of~$\nu'$.
Modes below 0.6\,THz in the crystal, or 0.4\,THz in the glass
contribute about 55\% to the total displacement;
90\% are reached only at about 2\,THz.
This means that the modes which are responsible for
the mean square displacement and the Debye-Waller factor
are not rigid body motions alone, 
but contain a significant contribution from 
intramolecular degrees of freedom.

\subsection{Debye-Limit and Sound Velocity} \label{Sound}

Assuming Debye behaviour at very low frequencies 
$g(\nu)=9\nu^2/\nu_D^3$, 
we can compare the neutron DOS
with the Debye frequency calculated 
from experimental density and sound velocities:
  \begin{equation}
       (2\,\pi \nu_{\rm D})^{-3}=\frac{V}{18 \pi^2 N}
        \left(\frac{1}{v_{\rm L}^3} +\frac{2}{v_{\rm T}^3} \right)
  \end{equation}
where $N$ is the number of molecules in volume~$V$.
For the sound velocities of the crystal,
we take the average $\langle v^{-3} \rangle^{-1/3}$ 
over the three lattice directions of our triple-axis experiment at 200\,K.
For the glass, we take literature data \cite{HiWa81} 
from Brillouin scattering at 220\,K.
The so-obtained $9/\nu_D^3$ are indicated by arrows in Fig.~3.

In both cases,
but in particular for the glass,
the neutron DOS extrapolates to a far higher Debye level than
expected from the sound velocities.
Part of the large discrepancy 
may be due to inaccurate estimates for
the sound velocity (see tables)
(for instance, the velocities from Brillouin scattering are based
on a temperature extrapolation of the refraction index)
or due to the circumstance of a broad tail of the resolution of IN6 
and a boson peak of OTP which 
is located at exceptionally low frequency.
But for the main part,
we must conclude that our neutron scattering experiment on the glass
simply does not reach the Debye regime.

In many other glass forming systems, 
similar discrepancies between DOS and sound velocities
are established as well \cite{BuPK88,GiRB93,WiBD98}, 
although in some substances a better accord is found
\cite{ZoAC95,WuPC95}.
Anyway, our data leave the possibility open
that the low-frequency spectrum of the glass and the crystal
contain non-harmonic, relaxational contributions,
as have been found recently by light scattering \cite{MoFM99,MoCL99}
or additional glassy excitation.

\subsection{Heat Capacity} \label{heat}

For a harmonic solid,
the heat capacity is given by the integral
\begin{equation}\label{cv}
   c_p(T) \simeq c_V(T) = N_{\rm at}R\int_0^\infty\!{\rm d}\nu\, g(\nu)
             \frac{(\beta/2)^2}{\sinh^2(\beta/2)}\,.
\end{equation} 
With a Debye DOS, this yields the well-known $c_p\propto T^3$.
Therefore, in Fig.~5 experimental data \cite{ChBe72} 
for the specific heat of glassy and polycrystalline OTP 
are plotted as $c_p/T^3$.
In this representation,
a boson peak at 0.35\,THz is expected to lead to a maximum at about 4\,K.

The lines which are calculated through (\ref{cv}) from the neutron DOS
agree for both glass and crystal in absolute units and 
over a broad temperature range with the measured data.
Similar accord has been reported for a number of other systems
\cite{CrBA94,BeCA96,TaRV98,WuPC95,BuPN86}.

At higher temperatures,
the heat capacities of crystalline and glassy OTP differ only little
({\it e.g.} at 200\,K: $c_{p}^{\rm cryst}= 182.8$\,J\,mol$^{-1}$\,K$^{-1}$ 
and $c_{p}^{\rm glass}= 186.1$\,J\,mol$^{-1}$\,K$^{-1}$ \cite{ChBe72}).
Towards high frequencies, the DOS becomes less sensitive
to the presence or absence of crystalline order 
(as suggested by the representation Fig.~3
for $\nu\gtrsim1\,$THz),
and remaining differences (clearly visible in Fig.~2)
are largely averaged out by the integral (\ref{cv}).

\section{Thermal effects} \label{sect4}

Figures~6--8
show the temperature evolution of vibrations 
in single crystal, polycrystal and glass.

In the single crystal, 
some phonons become softer on heating, others become stiffer,
depending on their direction, 
as exemplified in Fig.~6.
Broadening could not be observed because 
the linewidth remained always limited by the resolution of the spectrometer.

The temperature dependence of the different phonons is
in accord with results from Raman scattering on a polycrystal \cite{CrBA94}
where substantial positive and negative 
frequency shifts and broadening were observed already above 70\,K, 
indicating the presence of anharmonicities even at these low 
temperatures.
The opposite trends in the temperature evolution of different modes
ensure that the overall anharmonic effects are small, 
although individual modes are clearly anharmonic.

Through exceptionally large negative Gr\"uneisen parameters 
the softening of certain phonons may be related 
to anisotropic thermal expansion.
Negative expansion coefficient have indeed been found 
in crystalline OTP at much lower temperatures ($T<30$\,K) \cite{RaVB95}.

In the frequency distributions of the polycrystal
systematic temperature effects are detected
as well (Fig.~7a).
In Fig.~7b
we find a slight increase in the Debye limit ($\nu\to0$) of $g(\nu)/\nu^2$,
which may be explained by regular thermal expansion and softening ---
the kind of effects which can be accomodated in the harmonic theory of solids 
by admitting a temperature-dependent, ``quasiharmonic''
density of states $g(\nu;T)$.

In the glass the temperature effects are weak;
only around 2\,THz a systematic change in $g(\nu)$
may be recognised in Fig.~8a.
In the low-frequency DOS,
shown as $g(\nu)/\nu^2$ in Fig.~8b,
the temperature variation is stronger than in the polycrystalline counterpart.
The increase starts at about 160\,K, far below the glass transition.

The same anharmonicity of low-lying modes is also responsible
for the temperature dependence of the mean square displacement.
Fig.~4 shows that $\langle r^2(T) \rangle$ 
starts to increase faster than expected from (\ref{msd}) already
at about 140\,K.
These anharmonic contributions to $\langle r^2(T) \rangle$ amount to 
about 20\% at 200\,K for both phases, the crystal being slightly smaller.
Note, that around 140\,K deviations from the proportionality 
$\ln S(Q,\nu = 0) \propto T$ are also observed in 
coherent elastic scans on the BS instrument IN16.
The additional increase of $\langle r^2(T) \rangle$ 
above about 240\,K has been consistently interpreted 
as the onset of fast $\beta$ relaxation \cite{PeBF91}.
At higher temperatures,
in the presence of quasielastic scattering from relaxational modes,
the multi-phonon cross-section becomes ill-defined,
and the iterative determination of a DOS is no longer possible.

With a temperature-dependent DOS, the heat capacity
can be expressed as \cite{HuAl75}
\begin{equation} \label{anh-contr}
c_p(T) =
   N_{\rm at}R\int_0^\infty {\rm d}\nu\, g(\nu;T)
    \frac{(\beta/2)^2}{{\rm sinh}^2(\beta/2)} 
   \left[1-\left(\frac{\partial  \ln \nu }{\partial \ln T}
   \right)_p \right]
\end{equation}
where the second term arises explicitely from the shift of phonon modes.
The temperature derivative of the logarithmic moment
\begin{equation} \label{logmom}
\langle \ln \nu \rangle = \int_{0}^{\infty}\!{\rm d}\nu\, g(\nu) \ln \nu
\end{equation}
can then be taken as a direct measure for the degree of anharmonicity.
For Fig.~9, the integral~(\ref{logmom}) has been evaluated with 
an upper integration limit $\nu_{\rm g}=5\,$THz.
From the plot of $\langle \ln \nu \rangle$ versus $\ln T$ 
we estimate slopes 
${\partial \langle \ln \nu \rangle}/{\partial \ln T}$
of $-0.019$ for the polycrystal and $-0.012$ for the glass.
The stronger anharmonicity of the crystal 
can be traced back to the softening of high-frequency modes.

\section{Discussion} \label{sect5}

Our incoherent scattering experiments reconfirm
that it is possible to determine a meaningful DOS 
for a molecular system like OTP.
Cross-checks versus $\langle r^2(T) \rangle$ and $c_p(T)$ show
excellent accord 
if only the absolute scale of $g(\nu)$ is corrected for
multiple-scattering effects.

By coherent scattering on a single crystal low-lying phonon branches 
could be resolved. 
The low-frequency peaks in the DOS of the polycrystal
can be assigned to zone-boundary modes.
Optic phonons and hybridisation are found at rather low frequencies.
The excess of $g(\nu)$ of the glass over the crystal is
restricted to frequencies below about 0.5\,THz, 
the region where the crystal possesses mainly acoustic modes.
Towards higher frequency, 
the DOS is less structured in the glass than in the polycrystal,
but the overall spectral distribution is rather similar.

The main result of this communication concerns the strong thermal effects 
as they were manifested earlier close to the glass transition.
With increasing temperature,
the glass shows less anharmonicity than the polycrystal.
However, this may be partly due to a cancellation of opposite effects.
An example is provided by different phonon branches of the single crystal
for which positive, negative, and nearly vanishing temperature coefficients
are found.
In both, crystal and glass the anharmonic contributions to the mean square
displacement occur above $\sim150$\,K.
However, the anomalous increase in $\langle r^2(T) \rangle$
is much stronger in the glass, leading to the known glass
transition anomalies.

Finally, concering our previous analysis of quasielastic
scattering in the supercooled liquid OTP we can state:
Hybridisation and coupling between inter- and intramolecular modes
plays an important role for frequencies higher than 0.6\,THz.
As a consequence, the quasielastic scattering, which is confined to  
below 0.25\,THz, is clearly dominated by rigid-body motions and the analysis 
in terms relaxations in the framework of the mode coupling theory 
remains reasonable.

\section*{Acknowledgements}

We thank A.~Doerk (Institut f\"ur Physikalische Chemie, Mainz) 
for help in purifying several samples.
Financial support by BMBF under project numbers 03{\sc fu}4{\sc dor}5 
and 03{\sc pe}4{\sc tum}9 is gratefully acknowledged.

\section*{Appendix}

The determination of a density of states in absolute units
is complicated by the inevitable presence of multiple scattering.
In the inelastic spectrum of a glass, 
most multiple-scattering intensity
comes from elastic-inelastic scattering histories \cite{Sea75}.
This contribution is nearly isotropic, 
tends to smear out the characteristic $Q$ dependence of the phonon scattering,
and dominates at small angles where the single-scattering signal 
is expected to vanish as $Q^2$.

In our previous analysis of incoherent scattering from glassy OTP 
we performed a Monte-Carlo simulation for an ideal harmonic system
with given~$g(\nu)$ \cite{WuKB93}.
Here, we simply expand the scattering function 
$S(Q,\nu)=A(\nu) + Q^2 B(\nu) + \ldots$
which enables us to estimate the multiple scattering $A(\nu)$
by $Q^2 \to 0$ extrapolation
from the $S(2\theta(Q,\nu),\nu)$ spectra.
Results are shown in Fig.~10a.
For 200\,K, good accord with the Monte-Carlo simulation is found,
and measured spectra can be corrected by simply subtracting $A(\nu)$.
The interplay between multi-phonon scattering and multiple scattering 
limits this procedure to frequencies $\nu \lesssim 3$\,THz.

In the low-frequency region,
the subtraction of $A(\nu)$ leads to almost the same DOS 
as the {\it ad hoc} normalisation of Sect.~3.2.
This confirms our assignment of the 16 modes below the gap,
and it shows at the same time that multiple scattering is indeed the main
obstacle to a quantitative determination of a generalised density of
vibrational states.



\newpage


{\bf Figure captions:}\\
Figure 1: \\ 
Phonon dispersion relations of crystalline orthoterphenyl as
measured on the triple axis spectrometer IN12 at 200\,K.
The wave vectors are given in reciprocal lattice units.

Figure 2: \\
Renormalized density of vibrational states of glassy ($\square$) and 
crystalline OTP ($\blacksquare$) at 100\,K obtained from IN6.

Figure 3: \\
Density of vibrational states of glassy ($\square$) and 
crystalline OTP ($\blacksquare$) at 200\,K.
To emphasise the excess density of states of the glass over the crystal
$g(\nu)/\nu^2$ is shown.
The arrows mark the Debye limit $\nu \to 0$ calculated from sound
velocities and densities 
(at 220\,K for the glass and at 200\,K for the crystal).

Figure 4: \\
Mean square displacement $\langle r^2(T) \rangle$ of glassy OTP 
from elastic back-scattering (IN13, $\lozenge$)
and from time-of-flight spectra (IN6, plateau value of~$S(q,t)$, $\square$).
For comparison, IN6 data of polycrystalline OTP are also shown 
($\blacksquare$).
The lines are calculated from the densities of states at 100\,K 
of the glass (dashed) and the polycrystal (solid line).
The inset shows the relative value of $d=\langle r^2\rangle ^{1/2}$ 
at 100\,K for the crystal and the glass 
when the integral (\protect\ref{msd}) is restricted to modes with 
$0<\nu<\nu'$.

Figure 5: \\
Experimental heat capacity capacities $c_p/T^3$ of
glassy and crystalline OTP \protect\cite{ChBe72},
compared to $c_p/T^3$
calculated from the density of vibrational states $g(\nu)$ at 100\,K.

Figure 6: \\
Examples for the temperature dependence of phonons along different 
directions.
Hardening (left, note the negative energy scale), softening (middle) 
Our data leave the possibility open that the low-frequency spectrum of
the glass and the crystal contains non-harmonic, relaxational contributions.
and nearly no temperature
variation (right) is observed with increasing temperature.
The intensities are Bose corrected. The remaining differences are due
to the use of different set-ups (collimation).

Figure 7: \\
(a) Vibrational density of states of 
polycrystalline OTP for three different temperatures.
Note the significant temperature dependence of the DOS for all frequencies. ---
(b) To emphasise the temperature evolution at small energies,
the same data are plotted as $g(\nu)/\nu^2$.
The temperature dependence of the low-frequency modes becomes apparent.

Figure 8: \\
(a) Temperature dependence 
of the vibrational density of states of glassy OTP. ---
(b) The low frequency region of (a) plotted as $g(\nu)/\nu^2$.
Above 160\,K the density of low-frequency modes increases strongly.

Figure 9: \\
The logarithmic moment $\langle \ln \nu \rangle$ versus $\ln T$ 
of the density of states of glassy and crystalline OTP.
From the straight line,
one obtains the slope ${\partial \langle \ln \nu \rangle}/{\partial \ln T}$
which is a measure for the anharmonicity.

Figure 10: \\
(a) Comparison of the experimental data at two scattering angles with the 
multiple scattering contribution obtained by an $Q^2 \to 0$ extrapolation.
The MS nearly completely dominates the scattering signal at small angles. ---
(b) Comparison of the density of states obtained from the MS-corrected 
spectra with the {\it ad hoc} normalized DOS.

\clearpage
\newpage
\begin{table}
\caption{Sound velocities (in km/s)
for three phonon branches and different lattice directions,
obtained from linear fits to the low-$q$ limit
 of the measured phonon dispersions.}
\label{tab:1}  
\begin{tabular}{llllll}
\hline\noalign{\smallskip}
 &$v_{[100]}$ & $v_{[010]}$ & $v_{[001]}$ &
   $\langle v\rangle$ &${\langle v^{-3}\rangle}^{-1/3}$ \\
\noalign{\smallskip}\hline\noalign{\smallskip}
$v_{L}  $& 3.71  & 2.68  & 3.30 & 3.23 & 3.11 \\*[0.8em]
$v_{T_1}$& 1.82  & 1.52  & 1.97 && \\*[-.4em]
         &       &       &      & 1.75 &1.71 \\*[-.4em]
$v_{T_2}$& 1.67  & --    & --   &&\\
\noalign{\smallskip}
\hline
\end{tabular}
\end{table}

\begin{table}
\caption{Sound velocities in the glass as measured by Brillouin scattering,
using visible light or X-rays.}
\label{TsoundG}  
\begin{tabular}{lllll}
\hline\noalign{\smallskip}
  $T$ (K)&$v_{\rm L}$ (km/s)&$v_{\rm T}$ (km/s)&method&reference\\
\noalign{\smallskip}\hline\noalign{\smallskip}
220&2.94&1.37&light&\protect\cite{HiWa81}\\
223&2.63&&light&\protect\cite{MoCF98b}\\
200&2.70&&X-rays&\protect\cite{MoMR98}\\
\noalign{\smallskip}\hline
\end{tabular}
\end{table}


\begin{thebibliography}{}
\bibitem{x43}For a meaningful application of (\protect\ref{msd}), we have to
  assume the rigid-body limit (Sect.~3.1, Ref.~\protect\cite{WuKB93}).
  Therefore, $M$ has to be taken as the average atomic mass which in the case
  ot OTP is 7.2.
\end{thebibliography}
\end{document}